\documentstyle[aps,prb,epsf,floats]{revtex}

\begin{document}
\twocolumn[\hsize\textwidth\columnwidth\hsize\csname@twocolumnfalse%
\endcsname

\title {
Kondo effect of impurity moments in $d$-wave superconductors: \\
Quantum phase transition and spectral properties
}

\author{Matthias Vojta and Ralf Bulla}
\address{Theoretische Physik III, Elektronische Korrelationen und
Magnetismus, Institut f\"ur Physik,\\ Universit\"at Augsburg,
D-86135 Augsburg, Germany}
\date{August 10, 2001} 
\maketitle

\begin{abstract}
We discuss the dynamics of magnetic moments in $d$-wave
superconductors, in particular we focus on moments induced
by doping non-magnetic impurities into cuprates.
The interaction of such moments with the Bogoliubov
quasiparticles of the superconductors can be decribed
by variants of the pseudogap Kondo model, characterized by
a power-law density of states at the Fermi level.
The Numerical Renormalization Group technique is employed to investigate
this Kondo problem for realistic band structures and particle-hole
asymmetries, both at zero and finite temperatures.
In particular, we study the boundary quantum phase transition between the
local-moment and the asymmetric strong-coupling phases,
and argue that this transition has been observed in recent
nuclear magnetic resonance experiments.
We determine the spectral properties of both phases,
the location of the critical point as function of Kondo coupling and doping,
and discuss the quantum-critical cross-overs near
this phase transition.
In addition, the changes in the local density of states around
the impurity are calculated as function of temperature,
being relevant to scanning tunneling microscopy experiments.

\end{abstract}
\pacs{PACS numbers:}
]

\section{Introduction}
\label{sec:intro}

Impurities have proven to be a powerful probe for investigating
the bulk behavior of complex many-body systems.
In the field of high-temperature superconductivity,
a variety of phenomena have been
observed under doping with magnetic as well as non-magnetic
impurities:
suppression of the superconducting critical temperature $T_c$
and an increase of the residual in-plane resistivity
\cite{zntransport,kondotransport},
damping of collective magnetic excitations
(the so-called resonance mode \cite{keimer}),
possible pinning of stripes and vortices,
and impurity-related local bound states
seen in scanning tunneling microscopy (STM) \cite{seamuszn,ali,seamusni}.

A particularly interesting piece of physics is the magnetism
of impurities which are substituted for Cu ions.
Experiments have been performed with magnetic spin-1 (Ni) as
well as non-magnetic spin-0 (Zn, Li) impurities.
Whereas the spin-1 impurities naturally carry an on-site moment
which is expected to give rise to some kind of Kondo physics,
the behavior of non-magnetic impurities is more surprising.
A series of beautiful nuclear magnetic resonance (NMR) experiments
\cite{fink,alloul1,alloul2,mendels,orsay2,bobroff1,bobroff2,julien}
have clearly shown
that each impurity, despite having no on-site spin, induces a
local $S=1/2$ moment on the neighboring Cu ions at
intermediate energy scales.
In particular, the local susceptibility associated with Li ions
in YBa$_2$Cu$_3$O$_{6+x}$ (YBCO)
has been found to show Curie-Weiss-like behavior, with a strongly
doping dependent Weiss temperature which appears to vanish for
strongly underdoped samples --
this implies the existence of free moments in the underdoped regime
down to temperatures of order 1 K.
Microscopic perspectives on the formation of these
local impurity moments will be discussed in the next
section.

Accepting that non-magnetic impurities induce local moments,
these moments are expected to interact with the elementary
excitations of the bulk material \cite{phen}.
In this paper, we will be mainly concerned with a $d$-wave
superconducting bulk state, {\em i.e.}, with temperatures $T$
below the superconducting $T_c$:
the relevant low-energy excitations of the $d$-wave superconductor
are fermionic Bogoliubov quasiparticles and bosonic antiferromagnetic
spin fluctuations (as seen in neutron scattering);
quantum phase fluctuations which may become important
near a superconductor-insulator quantum phase transition
will not be considered here.
The interactions of impurity moments with fermionic and bosonic
degrees of freedom can be discussed separately,
as they lead to distinct phenomena associated with fermionic and
bosonic Kondo models.
Also, the energy scales of these two phenomena appear well separated,
as the fermionic Kondo temperature in the superconducting state is below 100 K
whereas the spin fluctuation energy scale is given by
the energy of the ``resonance mode'' \cite{keimer} being 40~meV
at optimal doping.

The interaction of an impurity spin with bosonic spin-1 fluctuations
of a quantum disordered antiferromagnet is described
by a ``bosonic Kondo model''~\cite{science,si,Sengupta}.
It has been studied recently in the context of
impurities in two-dimensional nearly-critical
antiferromagnets, where it leads to
a new (2+1)-dimensional boundary quantum field theory
with a number of interesting universal properties \cite{science}.
If the bulk state has a finite spin gap then the
impurity moment is not screened, {\em i.e.}, it contributes with
a Curie term to the impurity susceptibility.
However, the moment is spatially ``smeared'' over a length scale
given by the magnetic correlation length, {\em i.e.}, part of
the moment is carried by neighboring spins.
Another result of Ref.~\onlinecite{science} was that
a finite concentration of impurity moments leads to universal
damping of the collective spin-1 mode of the bulk magnet.
Together with recent neutron scattering experiments \cite{keimer}
this provides further evidence for $S=1/2$ {\em induced} local moments near
Zn sites in cuprate superconductors:
it has been argued\cite{science} that these moments are {\em required}
to explain the strong effects of a small concentration of Zn impurities
on the ``resonance peak'' in the spin dynamic structure factor.

\begin{figure}[!t]
\epsfxsize=2.5in
\centerline{\epsffile{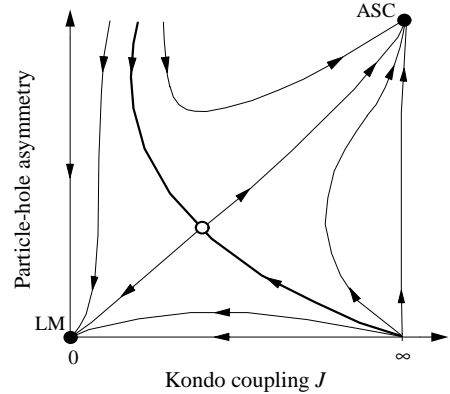}}
\caption{
Schematic zero-temperature
renormalization-group flow diagram \protect\cite{GBI}
for the pseudogap Kondo model with
power law density of states
$\rho(\epsilon) \sim |\epsilon|^r$ with $r>1/2$.
The horizontal axis denotes the Kondo coupling, the
vertical axis denotes particle-hole asymmetry
(parametrized, {\em e.g.}, by the potential scattering strength $U$
for a particle-hole symmetric host band structure).
The model has two stable fixed points (solid dots):
the weak-coupling local moment (LM) fixed point where
the impurity is essentially decoupled from the band, and
the asymmetric strong coupling (ASC) fixed point where
the impurity moment is fully screened.
The thick line represents the second-order phase transition
between the two zero-temperature phases,
the open dot is the critical fixed point.
Note that there is no screening in the particle-hole
symmetric case even for infinitely strong Kondo
coupling.
}
\label{fig:flow}
\end{figure}

In this paper, we will focus on the interaction of the
impurity moment with the {\em fermionic} bulk degrees of freedom,
namely the quasiparticles of the $d$-wave superconductor.
Importantly, the single-particle density of states (DOS) of the
Bogoliubov quasiparticles vanishes at the Fermi
energy, and so we expect a number of features quite
distinct from the usual Kondo effect in
metals \cite{hewson}.
The Kondo effect in systems where the host DOS
follows a power-law near the Fermi level,
$\rho(\epsilon) \sim |\epsilon|^r$ ($r>0$),
has been studied
quite extensively in the context of the pseudogap Kondo
model~\cite{withoff,cassa,ingersentsi,chen,bulla,ingersent,GBI,LMA,zhang,tolya2,OSPG}.
A number of studies~\cite{withoff,cassa,zhang,tolya2}
including the initial work by Withoff and Fradkin employed a slave-boson
large-$N$ technique~\cite{SB};
further progress and insight came from
numerical renormalization group (NRG) calculations \cite{chen,bulla,ingersent,GBI},
the local moment approach~\cite{LMA}, and a dynamic large-$N$ method \cite{OSPG}.
The general picture arising from these studies is that there exists
a zero-temperature (boundary) phase transition at a critical Kondo
coupling, $J_c$, below which the impurity spin is unscreened
even at lowest temperatures.
Also, the behavior depends sensitively on the presence or absence of
particle-hole symmetry: in the particle-hole symmetric case there
is no complete screening even for Kondo couplings $J>J_c$.
A comprehensive discussion of possible fixed points
and their thermodynamic properties has been given by Gonzalez-Buxton and
Ingersent \cite{GBI} based on the NRG approach.
Fig.~\ref{fig:flow} summarizes their findings for $r>1/2$, including the
case $r=1$ relevant
for $d$-wave superconductors
(see also Fig.~16 of Ref.~\onlinecite{GBI}).

Spectral properties of the pseudogap Kondo model
(for the relevant case of SU(2) spin symmetry)
have so far been investigated for the particle-hole symmetric
case only~\cite{bulla,LMA}.
However, interesting properties are to be expected especially
for the asymmetric case, considering
the results of static \cite{withoff,cassa,zhang} and dynamic \cite{OSPG}
large-$N$ approaches.

The study of spectral properties of Kondo impurities
in metals as well as in superconductors has been revived
due to recent advances in scanning tunneling
microscopy (STM).
Tunneling to clean surfaces of materials with bulk or
surface impurites has made it possible to directly
observe local spectral properties associated with impurities.
In particular, Kondo impurities on metal surfaces have been
found to give rise to a {\em dip} or Fano-like lineshape in
the tunneling DOS \cite{stmmetal,freddy}.
In contrast, in cuprate $d$-wave superconductors
{\em peak} structures in the local DOS close to the Fermi energy
have been found for both Zn and Ni
impurities ~\cite{seamuszn,ali,seamusni}
doped into ${\rm Bi}_2 {\rm Sr}_2 {\rm CaCu}_2 {\rm O}_{8+
\delta}$ (BSCCO).
Whereas the results for Ni appear to be well described
by a model with combined potential and static magnetic
(spin-dependent) scattering~\cite{martin},
the signal of Zn impurities -- a huge tunneling peak at
energies of 1--2 meV -- is more puzzling.
It is tempting to identify this peak with a quasi-bound state
in a purely potential scattering model\cite{martin,bala,tsu,zhuting,haas},
but such a state appears at low energies only for a
range of very large potential values
depending upon microscopic details \cite{martin,alan,ph}.
Furthermore, the spatial dependence of
the zero bias peak is surprising -- further discussion in
Sec.~\ref{sec:4site} --
and the observed spatially integrated spectrum
is asymmetric between positive and negative bias \cite{seamuszn},
while the potential model predicts approximate symmetry
\cite{flatte}.
Recently, it has been proposed \cite{tolya} that some of the
properties of the Zn resonance can be explained by taking
into account the Kondo spin dynamics of the magnetic moment
induced by Zn.
The screening of this moment by the Boboliubov quasiparticles
provides a natural low-energy scale (the Kondo temperature)
explaining the energetic location of the peak seen in STM.

The purpose of this paper is twofold:
On one hand we want to study the fermionic pseudogap Kondo
effect using realistic band structures for the cuprates.
We shall show that the transition
between a free moment and a screened moment in the
pseudogap Kondo model can be driven by
varying doping in the high-$T_c$ compounds -- this
provides an explanation for the strong doping
dependence of the Kondo (Weiss) temperature observed in
NMR \cite{bobroff1,bobroff2}.
On the other hand we want to examine a number of
properties of the pseudogap Kondo model
close to this transition and relate them
to experimental findings, in particular STM
measurements.

The remaining part of the paper is organized as follows.
In Sec.~\ref{sec:zn_models} we briefly discuss the issue
of properly modelling a non-magnetic impurity in cuprates --
we will motivate the model consisting of a Kondo and a potential
scattering term, and further discuss a number of theoretical
aspects in Sec.~\ref{sec:methods}.
In Sec.~\ref{sec:1site} we turn to the simplest model, namely
a point-like magnetic impurity with a realistic host density
of states (taking dispersion and gap data of actual cuprate
materials).
Using the numerical renormalization group method, we will
discuss static and dynamic properties of the two stable fixed
points as well as of the quantum critical regime.
We will determine the critical coupling constant
and the cross-over scale as function of doping, and
make contact with the transition seen in NMR experiments.
In Sec.~\ref{sec:4site} we extend the impurity modelling
to a spatially distributed magnetic moment and additional potential
scattering and present in particular
the resulting local conduction electron DOS
which is relevant for STM experiments.
A brief discussion of open issues will close the paper.

Parts of the discussion in Sec.~\ref{sec:zn_models} have
already been given in Refs.~\onlinecite{tolya} and \onlinecite{ssmv},
but we decided to include them here to keep the paper self-contained.
Readers mainly interested in the experimental consequences
for cuprates might skip the technical issues in
Sec.~\ref{sec:methods}, and have a look at
Sec.~\ref{sec:critpt} (especially Fig.~\ref{fig:tstardop}),
as well as Secs.~\ref{sec:tmatrix} and \ref{sec:4site}
which are related to STM observations.


\section{Effective models for non-magnetic impurities}
\label{sec:zn_models}

As NMR experiments show, non-magnetic impurities can induce
local moments in correlated hosts.
In this section we will briefly discuss theoretical perspectives
on this phenomenon.

To describe a non-magnetic impurity in a strongly correlated
superconducting system, two approaches appear possible.
(i) 
One starts with a basic model for the bulk system containing
correlation terms, {\em e.g.}, a one-band or three-band Hubbard
model with strong on-site interaction,
and describes the non-magnetic impurity by a potential
scattering term as suggested by its chemistry.
This approach is notoriously difficult, as a strongly correlated
model (which is hard to deal with even without impurity) has to be
treated including the impurity and in the low-temperature limit.
This task is almost impossible for current numerical techniques
such as Monte-Carlo or exact diagonalization~\cite{numnote},
and analytical approaches suffer from uncontrolled
approximations.
(ii) One separates the questions of moment formation and
interaction of the moment with the bulk system.
To describe the latter, one can employ {\em effective}
models for both the impurity and the bulk system.
The model for the bulk system has to contain only the
minimal ingredients to describe the desired
low-temperature physics -- in the present case this is a
$d$-wave BCS model which does not contain correlation
effects other than the Hartree-Fock pairing term.
The impurity term of the effective model has to account
for the correlation effects related to the introduction of
the impurity -- in the case of interest it has (at least)
to contain the magnetic moment and a scattering potential.

In this discussion and the rest of the paper, we will
adopt the approach (ii), see also
Ref.~\onlinecite{ssmv}.
Therefore, we will not attempt a rigorous derivation showing
the existence of the magnetic moment near the non-magnetic
impurity, but instead give some plausible arguments for
its appearance.
If we consider an undoped paramagnetic Mott insulator as reference
system at $T=0$, the moment formation is easily
seen to arise from breaking singlets \cite{fink,ssnr} by
removing spins.
These unpaired spins will remain in the vicinity of the
impurities if the host antiferromagnet has {\em confined}
spinons, {\em i.e.}, elementary spin-1 excitations.
In this picture each impurity can be understood as localized
``holon'' which binds the moment of a $S=1/2$ ``spinon''.
By continuity, the described mechanism is expected to be
effective also at small, but finite hole doping.
At larger doping, a related picture can be developed by analogy with the
theory\cite{hflm} for moment formation in the disordered metallic
state of Si:P---small variations in the potential combine with
strong local correlations to induce very localized spin
excitations.
Other theoretical perspectives\cite{nagaosa,pepin,fulde,fujimoto} on
local moment formation in the cuprates have also been given --
most of them are based on the related idea that strong potential
scatterers can form
quasi-bound states at the impurity sites near the Fermi level~\cite{pepin}.
Accounting for the strong local Coulomb repulsion, each
bound state will capture only a single electron, and the
low energy physics will again be described by an
impurity spin model.

It is evident that the magnetic properties associated with the
induced moment will strongly depend on the hole doping level.
The undoped limit, {\em i.e.}, the undoped paramagnetic Mott insulator,
has a free
$S=1/2$ moment near each impurity, while the strongly doped limit
(where electronic correlations are presumably weak) does not.
From continuity
one expects that the free $T=0$ moment will survive in the
superconducting state for a finite range of doping;
spin quantization suggests that the size of the moment
as measured by the Curie term in the susceptibility is
always $S=1/2$ independent of doping, {\em i.e.},
each moment will contribute a divergent Curie susceptibility
$\sim 1/4T$ even in the (underdoped) superconducting
state.
Then, a quantum critical point separates the weak and strong
doping limits.
On the strong doping side of this quantum critical point, the moment
is Kondo screened as $T \rightarrow 0$, and such a regime is
continuously connected to a regime, at higher doping, where the
moment does not even form at intermediate $T$.
We wish to argue that this quantum phase transition
is precisely the transition present in the pseudogap Kondo
model, and this claim is supported by quantitative calculations
in Sec.~\ref{sec:1site}.

We note that the pictures of the moment formation
quoted above do not give reliable information
about the parameters of a possible effective impurity
model.
NMR suggests that the magnetic moment is spatially
distributed among the Cu sites near the
impurity (but fluctuates as a single entity).
However, the precise microscopic form of
the coupling between the spin moment and the conduction
electrons is not known -- one has to keep in mind
that the moment is formed by a
particular (bound) state of the conduction electrons
near the impurity,
and it interacts with the other states
({\em i.e.} linear combinations) of the {\em same}
conduction electrons.
In absence of a precise knowledge of the appropriate
microscopic model we will employ some
simple Anderson- and Kondo-like models for the
impurities which are discussed in the following
section.


\section{Microscopic models and methods}
\label{sec:methods}

For the remainder of this paper we restrict our attention
to effective low-energy models consisting of a BCS-like
$d$-wave superconductor, a scattering potential, and a
magnetic moment,
${\cal H}={\cal H}_{\rm BCS} + {\cal H}_{\rm pot} + {\cal H}_{\rm mag}$.
In particular, we neglect the strong short-range magnetic
correlations of the bulk material.
As mentioned in the introduction, these low-energy
magnetic fluctuations couple to the impurity moment
leading to a bosonic Kondo model.
Due to the non-zero spin gap, the main effect
of this coupling is an additional spatial ``smearing'' of
each impurity moment, which can be absorbed in the effective
impurity model.

We assume that the cuprate superconductor can be described within
BCS theory, so we use
\begin{equation}
{\cal H}_{\rm BCS} = \sum_{\bf k} \Psi^{\dagger}_{\bf k} \left[
(\varepsilon_{\bf k} - \mu) \tau^z + \Delta_{\bf k} \tau^x \right]
\Psi_{\bf k}.
\label{hbcs}
\end{equation}
Here $\Psi_{\bf k} = (c_{{\bf k}\uparrow}, c_{-{\bf k}\downarrow}^{\dagger})$ is a
Nambu spinor at momentum ${\bf k}=(k_x, k_y)$ ($c_{{\bf k}\alpha}$ annihilates
an electron with spin $\alpha$ on a Cu 3d orbital),
$\tau^{x,y,z}$ are Pauli matrices in particle-hole space, and $\mu$ is
the chemical potential.
For the kinetic energy, $\varepsilon_{\bf k}$,
we will use a tight-binding form which includes
nearest neighbor hopping as well as longer-range hopping processes~\cite{disp},
while we assume a
$d$-wave form for the BCS pairing function $\Delta_{\bf k} =
(\Delta_0/2)
(\cos k_x - \cos k_y )$.

In the numerical calculations of this paper
we will neglect order parameter relaxation
due to the impurity.
This effect might play a role for
strong scattering potentials, in that it suppresses
the magnitude of the gap and leads to a finite zero-energy
DOS.
However, STM tunneling experiments \cite{seamuszn,ali}
show that the local change in the gap
size is rather small.
Furthermore, the induced zero-energy DOS will also
be small, and the resulting (finite) Kondo temperature
is exponentially suppressed.
So we expect results including order parameter
relaxation to be similar to ours below,
at least regarding the magnetic properties of the
impurity in the experimentally accessible temperature
range.


\subsection{Anderson vs. Kondo model}

An effective model for a (magnetic or non-magnetic)
impurity consists of a potential scattering term
and a magnetic term.
The potential scattering term is assumed to be localized at the
impurity site:
\begin{equation}
{\cal H}_{\rm pot} = U \sum_\sigma c_{0\sigma}^\dagger c_{0\sigma}
\,.
\label{hpot}
\end{equation}
The problem of a single scatterer can be solved exactly, with the
standard result for the $\Psi$ Green's function in Nambu notation:
\begin{eqnarray}
G({\bf r}&&,{\bf r}^{\prime},\omega) =
G^0 ({\bf r}-{\bf r}^{\prime},\omega)
- U G^0 ({\bf r}-{\bf r}_0,\omega_n) \nonumber \\
&&\times \tau^z [ 1 +
U G^0 ((0,0),\omega) \tau^z ]^{-1} G^0 ({\bf r}_0-{\bf
r}^{\prime},\omega);
\label{scatgf}
\end{eqnarray}
$G^0$ is the Green's function of the host $N_s G^0 ({\bf r}, \omega) =
\sum_{\bf k} e^{i {\bf k} \cdot {\bf r}} [\omega
- (\varepsilon_k -\mu) \tau^z - \Delta_k \tau^x ]^{-1}$,
and ${\bf r}_0=(0,0)$ is the scattering site.

The magnetic moment can be either coupled to a single site
(point-like impurity) or be spatially distributed.
For the point-like magnetic impurity coupled to a single site ${\bf r}=(0,0)$ of
a (metallic or superconducting) host the most general form
is a single-impurity Anderson model:
\begin{eqnarray}
{\cal H}_{\rm mag}
&=& \sum_\sigma V_0 (f_\sigma^\dagger c_{0\sigma} + h.c.) \nonumber \\
&+& \epsilon_f \sum_\sigma f_\sigma^\dagger f_\sigma + U_f n_{f\uparrow} n_{f\downarrow}
\,.
\label{hmag1}
\end{eqnarray}
In the so-called Kondo limit, $V_0\to \infty$, $\epsilon_f \to -\infty$, $U\to \infty$,
charge fluctuations on the impurity orbital are frozen,
and the Anderson model can be mapped onto a Kondo model (Schrieffer-Wolff transformation):
\begin{equation}
{\cal H}_{\rm mag} = J {\bf S} \cdot {\bf s}_0
\label{hmag2}
\end{equation}
where ${\bf s}_0 = N^{-1} \sum_{\bf kk'\alpha\beta} c^\dagger_{{\bf k}\alpha} \frac{1}{2} {\bf \sigma}_{\alpha\beta}
c_{{\bf k '}\beta}$ is the conduction band spin operator at the impurity
site ${\bf r}_0 = (0,0)$, $N$ the number of lattice sites,
and $J = 2 V_0^2 (1/|\epsilon_f| + 1/|U_f+\epsilon_f|)$.
For $U_f/2 \neq -\epsilon_f$ this mapping also introduces an additional potential scattering
term which can be absorbed in $U$ (\ref{hpot}).

Both models allow for a straightforward generalization to
spatially distributed impurities, which are, however, no longer equivalent!
The Anderson model
\begin{eqnarray}
{\cal H}_{\rm mag}
&=& \sum_{{\bf R}\sigma} V_{\bf R} (f_\sigma^\dagger c_{{\bf R}\sigma} + h.c.) \nonumber \\
&+& \epsilon_f \sum_\sigma f_\sigma^\dagger f_\sigma + U_f n_{f\uparrow} n_{f\downarrow}
\label{hmag3}
\end{eqnarray}
is easily seen to describe an Anderson impurity coupled to a single linear combination
of conduction electrons on the sites $\bf R$,
${\widetilde c} = \sum_{\bf R} V_{\bf R} c_{\bf R}/V$
where $V^2 = \sum_{\bf R} V_{\bf R}^2$.
In the Kondo limit, such a model maps onto a Kondo model with non-local Kondo
couplings.
In contrast, the straightforward generalization of the Kondo model (\ref{hmag2})
to an extended impurity reads
\begin{equation}
{\cal H}_{\rm mag} = \sum_{\bf R} J_{\bf R} {\bf S} \cdot {\bf s}_{\bf R} \,,
\label{hmag4}
\end{equation}
this model represents a multi-channel Kondo model \cite{NB,CZ}.
The impurity spin couples to all possible linear combinations
of the conduction electrons on the $\bf R$ sites.
In other words,
a spatially extended Kondo impurity generically
couples to different angular momentum channels of
the conduction electrons.

The relation between the models (\ref{hmag3}) and (\ref{hmag4})
has been recently discussed in Ref.~\onlinecite{ColemanMC}.
It has been argued that in an Anderson model like (\ref{hmag3})
the coupling to a {\em correlated} host opens new screening
channels, and the effective model will be a multi-channel Kondo
model (\ref{hmag4}).
The main idea is that conduction band correlations reduce
charge fluctuations in the host, and if an electron hops
onto the impurity site, it has to hop back preferentially onto
the same conduction band site where it came from.

We note that the different screening channels generated
by such a mechanism are certainly not equivalent,
and the low-$T$ behavior will be determined by a single-channel
fixed point. (However, this does not exclude that genuine
multi-channel physics may be realized at intermediate
temperatures.)
We will employ a Kondo model of the form (\ref{hmag4})
in Sec.~\ref{sec:4site} when discussing STM experiments
on Zn in BSCCO.


\subsection{Influence of superconducting correlations}

A Hamiltonian of the form ${\cal H}_{\rm BCS} + {\cal H}_{\rm mag}$
does not only describe a host with a pseudogap density of
states, it also contains superconducting correlations.
From diagrammatic pertubation theory it is clear
that the Kondo problem in
a normal-state system with a power-law DOS
({\em e.g.} a semi-metal) is not equivalent
to the Kondo problem in a superconductor with
the same DOS --
in other words, anomalous Green's function
contribute to the screening.

For the special case of a point-like Kondo impurity
in an $s$-wave superconductor, however,
it has been shown \cite{sakaisc} that the problem
with anomalous Green's functions
can be mapped onto a Kondo problem in
a normal-state system with modified gapped DOS,
containing an additional particle-hole asymmetry
({\em i.e.}, a potential scattering term),
but without anomalous Green's functions.

In $d$-wave superconductors such a mapping has not been
achieved so far, and the situation is less clear
because the anomalous Green's functions of the
$d$-wave superconductor
provide a coupling between different angular momentum
channels.
Therefore, it is expected that the Kondo problem in
a $d$-wave superconductor involves infinitely
many bands (angular momenta) even for a point-like
impurity.

Various approximate methods which have been employed
for solving the Kondo problem in a $d$-wave superconductor
handle the anomalous Green's functions differently.
Most calculations turn out to be insensitive to
superconducting correlations in the host material,
{\em i.e.}, they treat a semi-metal rather than a
superconductor.

For instance,
in the standard slave-boson approach, anomalous terms are
formally included, however, for a point-like
impurity in a $d$-wave superconductor their
contributions are easily seen to drop out.
The standard implementation of the numerical
renormalization group method uses the host density
of states as the only input, and anomalous terms
are ignored.
Based on the results for $s$-wave superconductors,
we nevertheless expect
that the results obtained by NRG are at least
qualitatively correct in a generic particle-hole
asymmetric situation, as the main influence of the
anomalous Green's functions is possibly an
additional particle-hole asymmetry.


\subsection{Numerical renormalization group method}

The NRG method has been developed by Wilson for the
investigation of the Kondo model~\cite{wilson}. Due to
the logarithmic discretization of the conduction band,
it is able to access arbitrarily low energy scales which
is essential also for the model discussed in this paper.
(However, due to the logarithmic discretization it is
not possible to resolve sharp structures at high energies.)
The NRG is a non-perturbative method as the
impurity site including the strong local Coulomb interaction
is treated exactly.

The NRG method has been generalized to a number of different
impurity models, such as the pseudogap Kondo and Anderson
models~\cite{chen,bulla,ingersent,GBI}.
For these models, the main modification is the mapping
of a non-constant conduction electron density of states onto
a semi-infinite chain form (see, {\em e.g.}, Ref.~\onlinecite{bulla}).

Recently the NRG method has been extended to the calculation
of dynamic properties at finite temperatures, we refer
the reader to Ref.~\onlinecite{BCV} for details.
We should mention here that information on spectral quantities for frequencies
$\omega \ll T$ cannot be determined from the NRG method; the resolution
is limited by the temperature and possible structures at
frequencies much smaller than the temperature are missed
by this approach.
However, this causes no problem when comparing with, {\em e.g.},
STM experiments, because the finite temperature in both
the sample and the tip introduces a thermal broadening,
such that structures on scales smaller than $T$ will not be
resolved.

The NRG calculations presented in this paper are mostly
performed with a discretization parameter $\Lambda=2$
and 600 or 800 levels.


\subsection{Slave-boson mean-field approximation}

The slave-boson mean-field approximation~\cite{SB} to
the Kondo impurity model is a large-$N$ approach,
{\em i.e.}, it becomes exact in the limit of
large spin degeneracy $N\to\infty$.
It is known that this method is not quantitatively
accurate for the physical case $N=2$, and has numerous
artifacts, both at finite temperatures in general
as well as near the quantum-critical point
in the pseudogap Kondo model.
However, it is expected to capture the qualitative physics
in a Kondo screened phase.

The method has been applied in a number of papers to the
pseudogap Kondo model~\cite{withoff,cassa,zhang,tolya2},
and we will mention some results below.


\section{Point-like magnetic impurities} 
\label{sec:1site}

In this section, we discuss general properties of the
asymmetric pseudogap Kondo model, in particular
the quantum phase transition between the local-moment and the
asymmetric-strong-coupling phase.
Since we are mainly interested in the universal
properties near this transition,
we restrict our attention here to
a $S=\frac{1}{2}$ impurity coupled to
a pseudogap band structure, and give
most results for general values of the pseudogap
exponent $1/2 < r \leq 1$
(some results are valid also for $0<r\leq 1/2$).
We will obtain numerical results using NRG, where
the DOS is the only host quantity entering the
calculation.
For technical purposes, namely a convenient calculation
of the impurity spectral function (or T matrix),
the impurity is modelled by an Anderson Hamiltonian (\ref{hmag1})
(instead of a Kondo Hamiltonian)
with $U_f = -2 \epsilon_f$, and parameters
$|U_f|, |\epsilon_f| \gg$ bandwidth which places the model
in the Kondo limit~\cite{kondorem}.

When specifying the results for the high-$T_c$ compounds,
we consider an impurity coupled locally to
a single site of the host $d$-wave superconductor.
To obtain quantitative estimates for characteristic energies,
we use tight-binding hopping parameters extracted
from normal-state angle-resolved photoemission (ARPES)
data \cite{disp} and sizes of the $d$-wave
gap as obtained from a number of
tunneling and ARPES experiments in the superconducting state
\cite{arpesgap,tunnelgap}.


\subsection{Zero-temperature phase transition and static properties}

As discussed in a number of papers over the last
years~\cite{withoff,cassa,ingersentsi,chen,bulla,ingersent,GBI,LMA,zhang,tolya2,OSPG}
the pseudogap Kondo model (with power-law DOS $\rho(\epsilon) \sim |\epsilon|^r$)
shows a phase transition as
a function of the Kondo coupling.
For pseudogap exponents $r>1/2$ there are
two stable zero-temperature fixed points \cite{GBI},
namely
(i) a local moment (LM) regime is reached for weak initial coupling --
here $J$ flows to zero, and the impurity is unscreened -- and
(ii) an asymmetric strong-coupling (ASC) regime which
is reached only for sufficiently large Kondo coupling and
particle-hole symmetry breaking.
The renormalization group flow is sketched in
Fig.~\ref{fig:flow}.

Particle-hole asymmetry is a relevant parameter in the
pseudogap Kondo problem, as seen in
Fig.~\ref{fig:flow}.
For a particle-hole symmetric host band structure, it can
be parametrized by the amount of potential scattering at
the impurity.
In the general case of an asymmetric host,
particle-hole asymmetry can still be cast into a {\em single}
number, in the case of a point-like Kondo impurity the real part
of the local host Green's function (after inclusion of potential
scattering) is a suitable choice.
We will see below that, {\em e.g.}, the location of the ``Kondo'' peak
will depend on the sign of this overall particle-hole asymmetry.
Concerning the cuprates, this overall sign cannot be
extracted from experiments or theory,
as it depends on details of the two-dimensional
band structure which place the van-Hove singularity either
below or above the Fermi level.
(Remarkably, this van-Hove singularity has never been observed
experimentally.)

For the following discussion,
it is useful to define susceptibilities describing the response
to external magnetic fields.
We allow for a space-dependent field $H_{\rm u}$ coupled to the host and
a local field $H_{\rm imp}$ at the impurity,
\begin{eqnarray}
{\cal H}_{\rm BCS} &\longrightarrow& {\cal H}_{\rm BCS} -
            \sum_{i\beta\gamma} H_{\rm u,\alpha}(i) c_{i\beta}^\dagger
            \sigma_{\beta\gamma}^\alpha c_{i\gamma}
\,, \nonumber \\
{\cal H}_{\rm mag} &\longrightarrow& {\cal H}_{\rm mag} - H_{\rm imp,\alpha} S_\alpha
\,.
\end{eqnarray}
With these definitions, a space-independent uniform field applied
to the whole system corresponds to $H_{\rm u}(i) = H_{\rm imp} = H$.
Response functions can be defined from derivatives of the free energy, $F = - T \ln Z$
($k_B=1$) as follows:
\begin{eqnarray}
\chi_{\rm{u},\rm{u}} (i,i^{\prime} ) &=& \frac{T}{3} \frac{\delta^2 \ln
Z}{\delta H_{\rm{u}\alpha} (i) \delta H_{\rm{u} \alpha} (i^{\prime} )}
\nonumber \\
\chi_{\rm{u},\rm{imp}} (i) &=& \frac{T}{3} \frac{\delta^2 \ln
Z}{\delta H_{\rm{u}\alpha} (i) \delta H_{\rm{imp}, \alpha}}
\nonumber \\
\chi_{\rm{imp},\rm{imp}} &=& \frac{T}{3} \frac{\delta^2 \ln
Z}{\delta H_{\rm{imp},\alpha} \delta H_{\rm{imp}, \alpha}}
\label{genchi}
\end{eqnarray}
From these quantities we can define various observables,
starting with the impurity contribution to the total susceptibility,
\begin{eqnarray}
\chi_{\rm imp}(T)
&=& \chi_{\rm imp,imp} + 2 \sum_i \chi_{\rm u,imp}(i)  \nonumber \\
&+& \sum_{i,i'} \left( \chi_{\rm u,u}(i,i') - \chi_{\rm u,u}^{\rm bulk}(i,i') \right)
\,,
\end{eqnarray}
where $\chi_{\rm u,u}^{\rm bulk}$ is the susceptibility of the bulk system in
absence of impurities.
The local impurity susceptibility, {\em i.e.}, the response of the
impurity spin to a local field, is given by
\begin{equation}
\chi_{\rm loc}(T) = \chi_{\rm imp,imp} \,,
\label{chiloc}
\end{equation}
which is equivalent to the zero-frequency impurity
spin autocorrelation function.
Importantly, NMR Knight shift experiments measure a different
quantity -- there a uniform field is applied to the whole sample,
and the local response is given by
\begin{equation}
\chi_{\rm NMR}(T) = \chi_{\rm imp,imp} +  \sum_i \chi_{\rm u,imp}(i) \,.
\label{chinmr}
\end{equation}
Note that sometimes this quantity is referred to as
``local susceptibility'', {\em e.g.}, in Ref.~\onlinecite{ingersentsi}.

After having defined the susceptibilities, we return
to the pseudogap Kondo problem.
We briefly recall the impurity thermodynamic properties
of the relevant stable fixed points (LM and ASC) --
these have been given, {\em e.g.}, in Ref.~\onlinecite{GBI}.
The local-moment fixed point has all characteristics
of a decoupled spin-1/2:
susceptibility $T \chi_{\rm imp} = 1/4$,
entropy $S_{\rm imp} = \ln 2$, and
specific heat $C_{\rm imp} = 0$.
At the asymmetric strong-coupling fixed point
the impurity spin is fully quenched:
$T \chi_{\rm imp} = 0$,
$S_{\rm imp} = 0$, and
$C_{\rm imp} = 0$.
Remarkably, the leading corrections~\cite{GBI} here are
$\Delta(T \chi_{\rm imp}), \Delta S_{\rm imp}, \Delta
C_{\rm imp} \propto T^{2r}$ for $0<r\leq 1$,
in other words, the impurity susceptibility
$\chi_{\rm imp}(T)$ vanishes as $T\to 0$ for $r>1/2$.

We note that there exists a third stable fixed point of the
pseudogap Kondo model, namely a symmetric strong coupling
fixed point. It is reached for pseudogap exponents
$r<1/2$ in the particle-hole
symmetric case for sufficiently strong Kondo coupling;
its thermodynamic properties are somewhat different from
the ASC behavior quoted above \cite{GBI}.
However, the real materials are generically particle-hole
asymmetric, and the physics of the particle-hole symmetric
strong-coupling fixed point is of no relevance here.


\subsection{Quantum-critical and cross-over behavior}

In this section, we focus on the properties of the impurity model
in the vicinity of the quantum-critical point separating the local
moment from the asymmetric strong coupling phase.
(Critical properties for the particle-hole symmetric pseudogap
model are different, and have been examined, {\em e.g.}, in
Refs.~\onlinecite{ingersentsi} and \onlinecite{bulla}.)

In the vicinity of the critical point, one can define an energy scale
$T^*$, which vanishes at the transition, and defines the cross-over
energy above which quantum-critical behavior is observed~\cite{book}.
If $T^*$ is much smaller than both the bandwidth and the
superconducting gap
we expect all observables to show
scaling behavior as function of $T/T^*$ and $\omega/T^*$.
Importantly, for the special case of $r=1$,
logarithmic corrections occur -- strictly speaking,
these corrections invalidate one-parameter scaling
behavior since a second (high-energy) scale
$\Lambda$ (of order bandwidth) enters in the
form of $\ln(T/\Lambda)$ or $\ln(\omega/\Lambda)$.
However, for most practical purposes the logarithmic
corrections are small, and in any case do not change
the leading power law behavior.
We have extracted the logarithmic corrections from
the NRG results for some observables, and they are
quoted below.

\begin{figure}
\epsfxsize=3.1in
\centerline{\epsffile{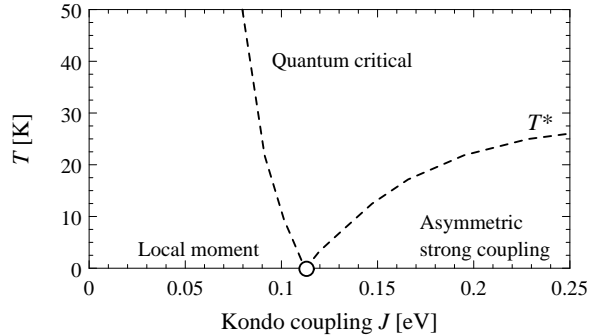}}
\caption{
Cross-over diagram of the pseudogap Kondo model
as function of the Kondo coupling $J$ and
temperature $T$, calculated with a point-like
impurity and a host band structure corresponding
to a cuprate superconductor at optimal doping.
The dot on the horizontal axis denotes the
quantum phase transition between LM and
ASC phases, the dashed lines are cross-overs,
indicating $T^\ast$.
Both quantum-critical and local-moment regimes
show a impurity susceptibility diverging as $1/T$,
whereas in the asymmetric strong coupling region the
impurity susceptibility vanishes $\propto T$.
As we will see in Sec.~\ref{sec:4site}, the
critical Kondo coupling becomes significantly
smaller for a spatially distributed impurity
moment.
}
\label{fig:tstar1}
\end{figure}

We start with the dependence of the cross-over
scale $T^\ast$ on the reduced coupling,
$j= (J-J_c)/J_c$, measuring the distance
from the critical point at $J_c$.
Our numerical results obtained by NRG
are consistent with
\begin{equation}
T^\ast \propto j^{1/r}
\end{equation}
for $1/2 < r \leq 1$ on both sides of the transition.
In Fig.~\ref{fig:tstar1}
we show numerical results for $T^\ast$
for a high-$T_c$ compound band structure at
optimal doping.
In Fig.~\ref{fig:tstar1} it is clearly
seen that even for Kondo couplings
well away from $J_c$, the cross-over scale $T^\ast$
is small (20 K), therefore impurity quantum-critical
properties may easily be observed in the cuprates
in a rather large temperature range without fine
tuning.
Parenthetically, we note that the cross-over scale $T^\ast$ is in general
defined only up to a prefactor of order unity;
in Fig.~\ref{fig:tstar1} the numerical value
of $T^\ast$ was taken from the location
of the maximum in the T matrix spectral density,
see below.

For $r=1$ the cross-over temperature
$T^\ast$ vanishes linearly with the distance to the
critical point (up to logarithmic corrections).
This is in agreement with earlier
NRG calculations~\cite{ingersent} as well as
with the dynamic large-$N$ approach~\cite{OSPG}.
However, it is at variance to the results obtained within the
slave-boson mean-field approximation \cite{cassa}
which predicts an essential singularity for $r=1$, {\em i.e.},
an exponentially vanishing scale $T^\ast$ near the
transition.
We believe that the NRG prediction of a linearly vanishing
$T^\ast$ is correct (as the NRG is reliable in capturing
asymptotic low-energy physics), and that the exponentially
vanishing scale is an artifact of the slave-boson
method.

\subsubsection{Susceptibility}

\begin{figure}[!t]
\epsfxsize=3.1in
\centerline{\epsffile{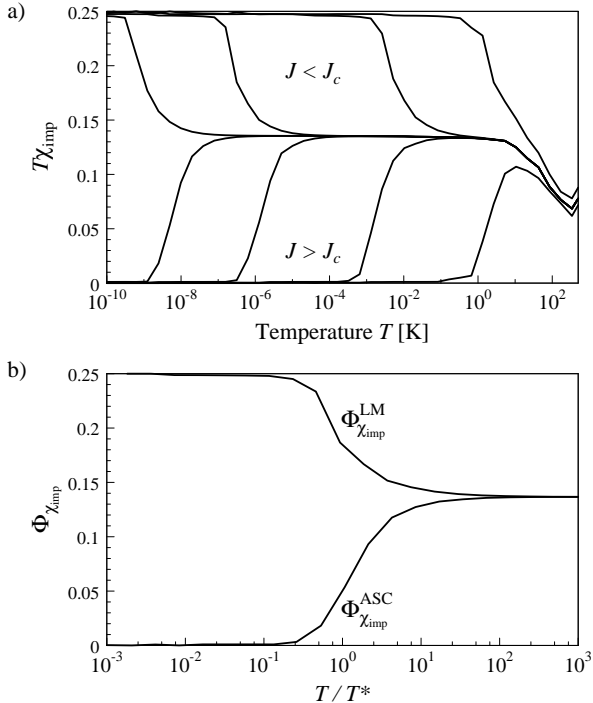}}
\caption{
a)
NRG results for the temperature dependence of $T\chi_{\rm imp}$,
for $r=1$, using a band structure corresponding to optimally doped
cuprates.
The different curves correspond to different values
of the Kondo coupling $J$ which are very close to
the critical coupling $J_c \approx 0.11 eV$.
b)
Scaling function $\Phi{\chi_{\rm imp}}(T/T^\ast)$ extracted from NRG
results, for $r=1$.
(Possible logarithmic corrections violating scaling are very
small here.)
The value of $T^\ast$ has been taken from the peak position
in the impurity spectral density, see Sec.~\ref{sec:tmatrix}.
}
\label{fig:tchi1}
\end{figure}

As a first observable we consider the impurity
susceptibility $\chi_{\rm imp}(T)$.
As a response function associated with a conserved quantity,
it cannot acquire an anomalous dimension \cite{conserved},
therefore we expect precisely at the critical point
a temperature dependence as
\begin{equation}
\chi_{\rm imp}(T, J\!=\!J_c) = \frac {\cal C} {T} \,.
\label{chiimpcrit}
\end{equation}
Here ${\cal C}$ is a universal number, depending only on $r$.
Interestingly, the behavior (\ref{chiimpcrit}) can be interpreted
as the Curie response of an irrational spin.
(${\cal C}=1/4$ would correspond to the usual Curie response of
a free spin 1/2.)
Numerical data for $T \chi_{\rm imp}$ are shown in
Fig.~\ref{fig:tchi1}.
In particular, close to $J_c$, there is a plateau in
$T\chi_{\rm imp}$, which allows to extract
the universal constant ${\cal C}(r=1) \approx 0.14$
in (\ref{chiimpcrit}).
This number is somewhat smaller than the value 0.164
quoted by Gonzalez-Buxton and Ingersent (Fig.~14 of Ref.~\onlinecite{GBI}),
this might be due to discretization or cut-off effects in
the numerics
(we have not attempted a careful finite-size analysis here).

The local susceptibility (\ref{chiloc}) is not protected by a conservation law,
and can acquire an anomalous exponent $\eta_{\rm loc}$.
This implies a power law
\begin{equation}
\chi_{\rm loc}(T, J\!=\!J_c) \propto T^{\eta_{\rm loc}-1}
\label{chiloccrit}
\end{equation}
at the critical point.

For both finite $T$ and finite $(J-J_c)$, the low-energy behavior
is completely determined by the cross-over energy
scale $T^\ast$ and the temperature $T$ itself
(up to logarithmic corrections for $r=1$).
Then, the susceptibilities can be described by universal
cross-over functions,
\begin{eqnarray}
\chi_{\rm imp}(T)
&=& \frac {1} {T} \, \Phi_{\chi_{\rm imp}} \!\left(\frac{T}{T^\ast}\right)
\nonumber \\
&=& \frac {T^{2r-1}} {T^{\ast 2r}} \, \widetilde{\Phi}_{\chi_{\rm imp}} \!\left(\frac{T}{T^\ast}\right)
\label{chiimpcross}
\end{eqnarray}
with $\widetilde{\Phi}{\chi_{\rm imp}}(x) = \Phi{\chi_{\rm imp}}(x) / x^{2r}$, and
\begin{eqnarray}
\chi_{\rm loc}(T)
&=& \frac {\cal{B}} {T^{1-\eta_{\rm loc}}} \, \Phi_{\chi_{\rm loc}} \!\left(\frac{T}{T^\ast}\right)
\nonumber \\
&=& \frac {{\cal B} T^{\ast \eta_{\rm loc}}} {T} \, \bar{\Phi}_{\chi_{\rm loc}} \! \left(\frac{T}{T^\ast}\right)
\label{chiloccross}
\end{eqnarray}
with $\bar{\Phi}{\chi_{\rm loc}}(x) = x^{\eta_{\rm loc}} \Phi{\chi_{\rm loc}}(x)$,
and $\cal B$ an amplitude prefactor.
The equations (\ref{chiimpcross}) and (\ref{chiloccross})
define universal scaling functions $\Phi$ of the reduced
temperature $T/T^\ast$; the $\Phi$ are, of course, different for
different $r$ and for both sides of the quantum phase transition, {\em i.e.}, we
have to distinguish $\Phi^{\rm LM}$ and $\Phi^{\rm ASC}$.
The scaling functions $\Phi{\chi_{\rm imp}}$ for $r=1$ are shown in
Fig.~\ref{fig:tchi1}b -- here our present numerics is not
accurate enough to observe possible logarithmic
corrections to scaling.
Also, a reliable fitting of $\chi_{\rm imp}$ to a Curie-Weiss law
for a temperature range above $T^\ast$ requires a finer discretization in
the NRG procedure, and has not been attempted here.
It is clear that the Weiss temperature will be approximately given
by the cross-over temperature $T^\ast$ (defined by the T matrix peak),
but we cannot give a reliable estimate for the ratio of the two.

We briefly discuss the asymptotics of the scaling functions $\Phi_{\chi}$.
Spin quantization requires $\Phi_{\chi_{\rm imp}}^{\rm LM}(0)=1/4$ and
$\Phi_{\chi_{\rm imp}}^{\rm ASC}(0)=0$,
therefore the zero-temperature limit of $T\chi_{\rm imp}$ is fixed to 1/4
throughout the whole local-moment phase (and to 0 in the whole ASC phase).
The universal Curie response at the critical point
(\ref{chiimpcrit}) implies
$\Phi_{\chi_{\rm imp}}^{\rm ASC}(\infty) = \Phi_{\chi_{\rm imp}}^{\rm LM}(\infty)={\cal C}$.
From $\chi_{\rm imp}(T\to 0)\propto T^{2r-1}$ in the ASC phase we
deduce $\Phi_{\chi_{\rm imp}}^{\rm ASC}(x) \propto x^{2r}$ for small $x$.
This immediately gives
$\chi_{\rm imp}(T=0)\propto T^{2r-1}/T^{\ast 2r} \propto T^{2r-1}/(J-J_c)^2$
for $J>J_c$.
However, the real superconductor will have a small residual DOS
at the impurity site due to order parameter relaxation,
this will will lead to a finite impurity susceptibility in the
zero-temperature limit for $J>J_c$.
Finally, NRG data \cite{GBI} indicate that $\chi_{\rm loc}(T\to 0) \propto 1/T$
in the LM regime;
this gives us $\Phi_{\chi_{\rm loc}}^{\rm LM}(x) \propto x^{-\eta_{\rm loc}}$,
and therefore $T\chi_{\rm loc}(T\to 0)\propto T^{ \ast \eta_{\rm loc}}
\propto (J-J_c)^{\eta_{\rm loc}/r}$ in the LM phase.
Our preliminary NRG calculations of $T \chi_{\rm loc}$ allowed to
extract the value of $\eta_{\rm loc}$, we found
$\eta_{\rm loc} \approx 0.05$ for $r=1$.

At this point a few remarks about experimental susceptibility
measurements and the possibility to observe the described
quantum-critical behavior are in order.
As detailed above, the only $T$ dependence with a non-trivial
power law can be expected in the local susceptibility
$\chi_{\rm loc}(T)$. However, a direct measurement of
$\chi_{\rm loc}$ requires a {\it local} field, and is
perhaps only possible using muon spin resonance.
In contrast,
commonly used NMR techniques probe $\chi_{\rm NMR}$ (\ref{chinmr}),
and this quantity shows either a Curie law or saturates
to a constant at low $T$.
Measurements of the {\it total} susceptibility
in the local-moment regime should observe Curie behavior
with the ``full'' prefactor
corresponding to spin-1/2 per impurity, {\em i.e.},
$\chi_{\rm imp} = N_{\rm imp}/4T$, where $N_{\rm imp}$
is the number of impurities.
Such measurements have been performed on Zn-doped YBCO \cite{mendels},
and the results indicate a somewhat smaller value of the moment
per impurity, which on the other hand depends only weakly
on doping.
Two explanations for the deviation from the $S=1/2$ moment
appear to be possible:
Either there are subtle cancellation effects involved in the
SQUID susceptibility measurements (one also has to keep in mind
that the subtraction of the bulk susceptibility leads to rather
large uncertainties in $\chi_{\rm imp}$),
or the observed reduction of the moment reflects the
quantum-critical behavior (\ref{chiimpcrit}).
In the latter case, one should expect a cross-over
between the Curie law (\ref{chiimpcrit}) to
a Curie law with the ``full'' prefactor at
low enough temperatures, {\em i.e.}, for $T \ll T^\ast$,
on the local-moment side of the Kondo phase transition.
Experimentally, this would require measuring the
total susceptibility well below the superconducting $T_c$
on the underdoped side, and has not been performed
so far.

\subsubsection{T matrix}
\label{sec:tmatrix}

We now turn to dynamic properties associated with the
Kondo impurity, in particular we are interested
in the conduction electron T matrix.
The knowledge of the T matrix is important for
calculating the local density of states around the impurity,
as observed in STM experiments.
In the Anderson impurity model, the T matrix is connected
to the impurity Green's function according to \cite{hewson}
\begin{eqnarray}
T(\omega) = V^2 G_{\rm imp} (\omega)
\,.
\end{eqnarray}
We can define the T matrix spectral density as
$\rho_{\rm T}(\omega) = - {\rm Im}\,T(\omega) / \pi$.

The behavior of the impurity spectral function has
been studied in the particle-hole symmetric
pseudogap Anderson model in Ref.~\onlinecite{bulla}.
In this case the strong coupling behavior is different
from the present asymmetric model, but the asymptotic
local-moment behavior is identical -- this is also
seen in Fig.~\ref{fig:flow}, where the particle-hole
asymmetry flows to zero in the LM phase.
For the LM regime it was found \cite{bulla} that the impurity
spectral density vanishes at the Fermi level as
$|\omega|^{r}$, this could be verified in the
present calculations.
In the asymmetric strong-coupling regime
we find similar behavior,
${\rm Im}\,G_{\rm imp}(\omega) \propto \rho_{\rm T}(\omega) \propto |\omega|^{r}$.
Interestingly, the prefactor of this power law is equal
for both positive and negative frequencies,
meaning that the spectral asymmetry disappears in the scaling
limit of ${\rm Im}\,G_{\rm imp}$ and $\rho_{\rm T}$.

Of particular interest is the behavior of $\rho_{\rm T}$
in the quantum-critical region.
From general scaling arguments, we expect a power law
$\rho_{\rm T} \propto \omega^{\eta_{\rm T}-1}$, where $\eta_{\rm T}$
is the anomalous exponent.
Our NRG results are consistent with such a power law, with
$\eta_{\rm T}=1-r$. So we have
\begin{eqnarray}
\rho_{\rm T}(\omega) &\propto& \frac{1}{\omega^r} \: ~~~~~~~~~~~~~~(r < 1) \,, \nonumber \\
\rho_{\rm T}(\omega) &\propto& \frac{1}{\omega \ln^2(\Lambda/\omega)} ~~~(r=1)
\,.
\end{eqnarray}
Similar to the LM and ASC behavior, any asymmetry drops out in the
scaling limit of $\rho_{\rm T}$, and the fixed-point spectral density
is particle-hole symmetric.

\begin{figure}[!t]
\epsfxsize=3.2in
\centerline{\epsffile{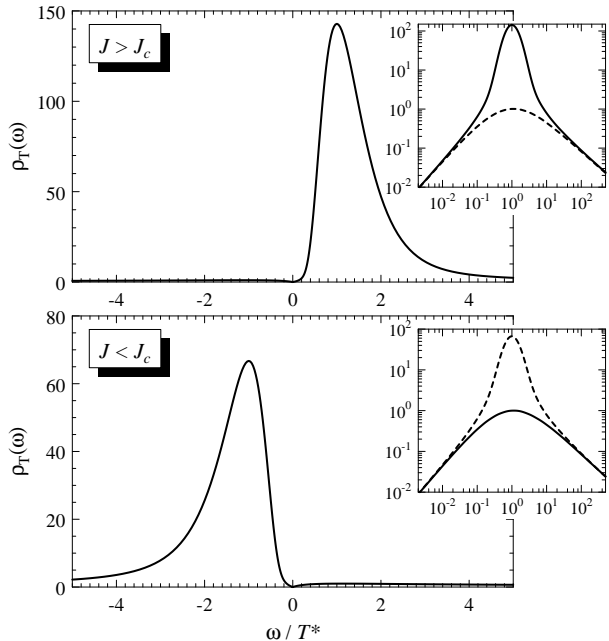}}
\caption{
Zero-temperature spectral densities $\rho_{\rm T}$
for the conduction electron T-matrix at $r=1$,
obtained by NRG.
Upper panel: $J>J_c$, cross-over from the ASC behavior
at low energies to quantum criticality
at high energies.
Lower panel: $J<J_c$, {\em i.e.}, cross-over from the LM
to quantum critical behavior.
For comparison, the vertical axis has been normalized
to the maximum value of $\rho_{\rm T}$ on the ``non-peak
side''.
Insets: Same on log-log scale, showing both
$\bar{\omega}>0$ (solid) and $\bar{\omega}<0$ (dashed).
Note that $\bar{\omega}>0$ and $\bar{\omega}<0$ are
interchanged when the overall particle-hole asymmetry
changes sign.
If we ignore logarithmic corrections, the functions
shown are equivalent to the imaginary part of the
scaling function $\Phi_{\rm T}$.
}
\label{fig:gimpteq0}
\end{figure}

As done for the susceptibilities, the cross-over between the quantum-critical
and the LM or ASC behavior in the T matrix can be described by a universal
scaling function:
\begin{equation}
T(\omega) = \frac{{\cal A}}{T^{\ast r}} \,
\Phi_{\rm T} \! \left(
\frac{\omega}{T^*}, \frac{T}{T^*} \right)
\,,
\label{gscale}
\end{equation}
where ${\cal A}$ is an amplitude prefactor,
and $\Phi_{\rm T}$ is a universal scaling
function (for the particular value of $r$ and for each side
of the transition).
For $r=1$ we have to keep in mind that logarithmic corrections
spoil scaling, and a form like (\ref{gscale}) is only approximately
valid.

Numerical results for the $T=0$ impurity Green's function
for $r=1$ are shown in Fig.~\ref{fig:gimpteq0}.
They correspond to parameter values very close to the
transition, and represent the scaling functions
$\Phi_{\rm T}$ if we neglect logarithmic
corrections.
For $\bar{\omega} = \omega/T^* \ll 1$ we have ASC behavior with
${\rm Im} \Phi_{\rm T} \sim |\bar{\omega}|$,
for $\bar{\omega} \gg 1$ the spectral density follows the
quantum-critical ``power law'' $|\bar{\omega}|^{-1} \ln^{-2}(\bar{\Lambda}/\bar{\omega})$
as predicted above, with
particle-hole symmetry in both limits.
However, in the cross-over region the asymmetry leads to a large
peak for one sign of $\bar{\omega}$
(depending on the sign of the overall particle-hole asymmetry of the
model) --
this is true for both sides of the quantum phase transition.

In other words,
the Kondo peak known from the metallic host Kondo model
which is located at the Fermi level (in the scaling limit)
is replaced by a peak at a {\em finite} energy -- this
energy actually corresponds to the cross-over temperature $T^\ast$
between the quantum-critical and the LM or ASC behavior.
For the cross-over from critical to ASC behavior, one can identify
$T^\ast$ with the ``Kondo'' temperature of the problem, since
a cross-over from a Curie divergence to a constant is seen
in $\chi_{\rm imp}(T)$ around $T^\ast$.
In this paper, we have used the location of the peak
in $\rho_{\rm T}(\omega)$ (Fig.~\ref{fig:gimpteq0})
as {\em definition} of $T^\ast$, as it is the numerically most
convenient and precise criterion.

\begin{figure}
\epsfxsize=3.3in
\centerline{\epsffile{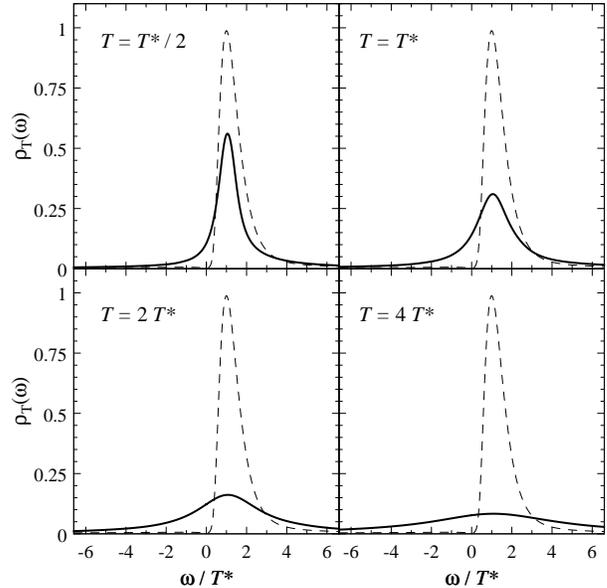}}
\caption{
Same as Fig.~\protect\ref{fig:gimpteq0},
but now for finite temperatures and $J>J_c$.
The four panels show the evolution of the cross-over peak for
temperatures $T = T^\ast/2,T^\ast,2T^\ast, 4T^\ast$,
the dashed line is the $T=0$ result.
The intensity has been normalized to the height of
the $T=0$ peak.
The ``Kondo'' peak is broadened and
slowly looses weight for temperatures above
$T^\ast$.
}
\label{fig:gimptg0}
\end{figure}

We note that very similar results for the T matrix have been found in a
recent dynamic large-$N$ study of a multi-channel pseudogap
Kondo model \cite{OSPG}. There it was possible to analytically
determine the low-energy behavior of the quantities of interest,
both at $T=0$ and finite $T$.
Related observations have also been made within slave-boson
mean-field calculations \cite{cassa,zhang}.

\begin{figure}
\epsfxsize=3in
\centerline{\epsffile{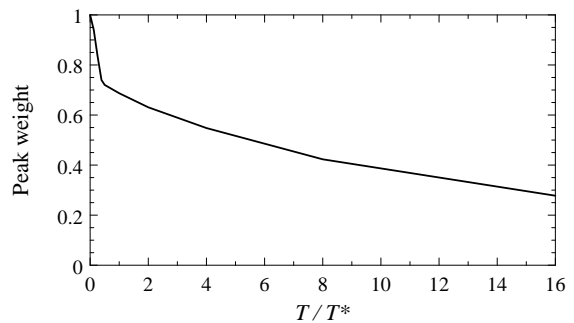}}
\caption{
Weight of the cross-over peak as shown in
Fig.~\protect\ref{fig:gimptg0}, integrated from
$\omega=-10 T^\ast$ to 10 $T^\ast$, and
normalized to the peak weight at $T=0$.
}
\label{fig:peakweight}
\end{figure}

We have also calculated the finite-temperature behavior
of the impurity T matrix by NRG,
for a fixed (temperature-independent) host density of
states.
It is worth emphasizing that the finite-$T$ NRG method
is a unique tool for this task, as the commonly used slave-boson
method shows a spurious finite-$T$ transition.
A series of cross-over functions for the
spectral intensity is shown in Fig.~\ref{fig:gimptg0}.
As expected, the peak is broadened, and
looses spectral weight for $T \gg T^\ast$,
whereas the peak position is nearly unaffected by changing $T$.
The evolution of the weight under the peak (integrated from
$\omega=-10 T^\ast$ to 10 $T^\ast$) is shown
in Fig.~\ref{fig:peakweight};
similar to the usual Kondo effect the weight loss
occurs rather slowly, {\em e.g.}, the weight is reduced
below 50 \% of its $T=0$ value only for $T > 5 T^\ast$.

The results in Figs.~\ref{fig:gimpteq0} and \ref{fig:gimptg0}
are possibly of relevance for the STM experiments
done on Zn-doped BSCCO \cite{seamuszn,ali} -- the large peak
in the differential conductance close to zero bias
correspond to the Kondo peak arising from the screening of the
Zn-induced moment.
We will discuss this issue further in Sec.~\ref{sec:4site}.

\subsection{Location of the critical point}
\label{sec:critpt}

So far we have discussed magnetic screening in the pseudogap
Kondo model for a fixed host density of states.
Motivated by NMR experiments on Zn- and Li-doped cuprates
we now consider the dependence of the pseudogap Kondo effect
on the hole doping level, {\em i.e.}, with varying concentration
of mobile carriers in the CuO$_2$ planes.
Experimentally, a strong dependence of the Kondo temperature
on the doping level has been observed~\cite{bobroff1,bobroff2},
in particular $T_K$ appears to vanish in strongly
underdoped samples.

As explained in the introduction, a (boundary) transition
between an unscreened moment for small doping and a
screened (or even absent) moment for large doping
can be anticipated on theoretical grounds.
We propose that this transition corresponds to the
quantum phase transition discussed above in the pseudogap
Kondo model (see also Ref.~\onlinecite{ssmv}),
and we shall show numerical data indeed supporting
that the transition in the pseudogap Kondo model can be driven
by changing the doping level.

Doping in the cuprates ({\em e.g.}, adding or removing oxygen in YCBO)
is known to have a number of important effects,
the main one of course being the change of the carrier
concentration.
For the pseudogap Kondo physics, the local density of
states is of particular importance.
If we stick to the simple, doping-independent, tight-binding
band structure used so far, then doping has two effects:
(i) the change in the chemical potential, influencing
carrier concentration and band asymmetry,
(ii) the change in the magnitude of the $d$-wave gap
as seen, {\em e.g.}, in tunneling and photoemission
experiments.
As we will see, point (ii) is most important for
the pseudogap Kondo physics, since a change
in gap size immediately influences the low-energy
part of the host DOS.

\begin{figure}
\epsfxsize=3in
\centerline{\epsffile{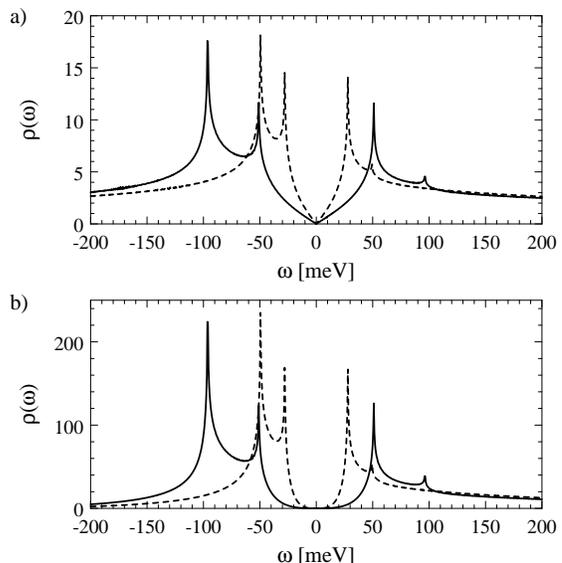}}
\caption{
Low-energy part of the input DOS used for the NRG calculations,
here without potential scattering, $U=0$.
Solid lines: 10\% underdoped.
Dashed lines: 19\% overdoped.
a) DOS for a single-site impurity, being equivalent to the local DOS
of the $d$-wave superconductor.
b) Effective DOS for a four-site impurity as discussed in
Sec.~\protect\ref{sec:4site}.
}
\label{fig:dos}
\end{figure}

For BSCCO it is established from both ARPES \cite{arpesgap}
and tunneling \cite{tunnelgap} measurements that the magnitude of the
superconducting gap (as well as the normal state pseudogap)
changes considerably with doping:
in contrast to $T_c$ the gap increases in the underdoped regime,
there taking up to twice of its value at optimal doping.
For YBCO the situation is less clear since high-quality
tunneling or ARPES data are rare~\cite{yeh}.
At least, thermodynamic measurements indicate an increase of
the pseudogap temperature with decreasing doping also in YBCO,
therefore a decrease of low-energy spectral weight similar
to BSCCO is likely.
In the following, we will use the doping-dependent gap values
as measured in BSCCO \cite{arpesgap,tunnelgap}, but we expect
the results to reproduce the correct trend for most cuprates.
The precise gap values used are
$\Delta_0 =$ 57, 38, 22 meV for doping $\delta =$ 10, 15, 20 \%.
The host band structure entering the Hamiltonian (\ref{hbcs})
is assumed to have a doping-independent
tight-binding form including longer-range
hopping terms,
we have performed calculations with
the three parameter sets quoted in Table I of Ref.~\onlinecite{disp}
as well as a simpler dispersion with
$t = -0.15$ eV, $t'=-t/4$, $t''=t/12$,
with qualitatively similar results.
The employed densities of states are shown in
Fig.~\ref{fig:dos}a for underdoped and overdoped
hosts;
Fig.~\ref{fig:dos}b shows the effective DOS seen
by a four-site impurity, see Sec.~\ref{sec:4site}.

Of course one has to keep in mind that the evolution
from the optimally doped $d$-wave superconductor to an
insulator in the underdoped limit involves Mott physics
which is not included in the simple one-particle description
employed here.
Therefore, the depletion of low-energy spectral weight,
here modelled by an increasing $d$-wave gap, should be
viewed as a phenomelogical account for strong correlation
effects.

\begin{figure}
\epsfxsize=3in
\centerline{\epsffile{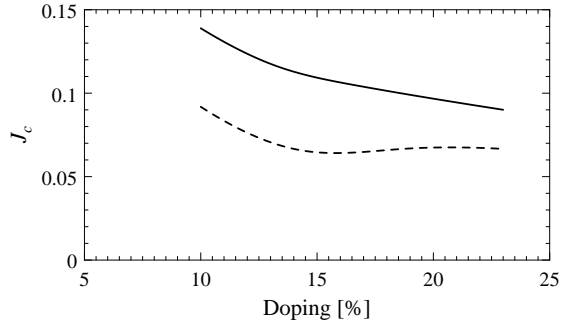}}
\caption{
Critical value of the Kondo coupling $J_c$ vs. doping,
extracted from NRG calculations for a cuprate host DOS, with
a doping-dependent superconducting gap.
Solid: single-site impurity.
Dashed: four-site impurity (see Sec.~\protect\ref{sec:4site}).
}
\label{fig:jc}
\end{figure}

NRG results for the doping dependence of the critical $J_c$
are shown in Fig.~\ref{fig:jc}, for both
a single-site impurity and a spatially distributed
four-site impurity (see Sec.~\ref{sec:4site}).
The data show a 30 to 50\% variation of the critical $J$,
induced by the size change of the superconducting gap.
For comparison we have also calculated the
critical $J$ with doping-independent gap -- the effect
of the change in asymmetry is very small due to the
presence of a finite asymmetry already at zero doping
arising from hopping processes beyond nearest neighbors.
In addition, we have checked that the change in the
{\em form} of the superconducting gap as reported~\cite{arpesgap}
for BSCCO, namely
the deviation from the $(\cos k_x - \cos k_y)$ form
for underdoped samples, has
only a weak effect on the critical $J$.
We note that additional potential scattering terms
of course modify the $J_c$ values, but we have
verified that for moderate values of the bare scattering
potential $U$ the trend that $J_c$ increases for strong
under-doping is preserved.

\begin{figure}
\epsfxsize=3in
\centerline{\epsffile{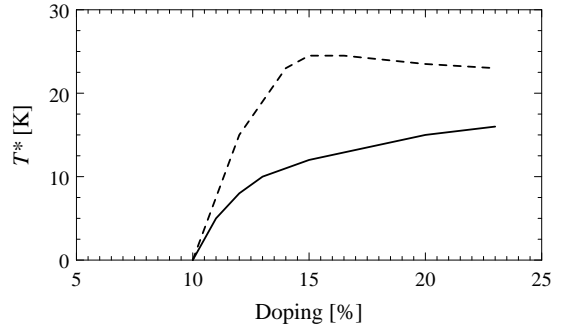}}
\caption{
Doping dependence of $T^\ast$ describing the cross-over
from quantum-critical to asymmetric strong-coupling behavior,
calculated with the band structures and gap values
used for Fig.~\protect\ref{fig:jc}.
For the impurity susceptibility the displayed $T^\ast$ corresponds
to the cross-over from Curie-like behavior to a constant.
Solid: single-site impurity with $J=0.14$ eV.
Dashed: four-site impurity with $J=0.09$ eV.
}
\label{fig:tstardop}
\end{figure}

The main message of Fig.~\ref{fig:jc} is that
$J_c$ acquires a significant doping dependence,
and has a value in the range of magnetic couplings which
are expected in the cuprates.
Of course, the precise form of the microscopic
Hamiltonian describing the induced moment is not
known, and therefore no estimate of the coupling value
can be given.
If we just assume a doping-independent Kondo coupling
of a certain size,
then we can obtain the values of the cross-over temperature
$T^\ast$ from our NRG calculations, and sample data are
shown in Fig.~\ref{fig:tstardop}.
The important result is that a {\em smooth} variation of
the input parameters can lead to
a {\em strongly} doping-dependent cross-over (``Kondo'')
temperature arising from the vicinity to the boundary quantum phase
transition.
Therefore, the experimentally observed strong doping
dependence of the NMR Weiss temperature~\cite{bobroff1,bobroff2}
is not necessarily an indication of a drastic change in the
host electronic structure.

We close this section with a remark on the analytical
understanding of the doping dependence of $J_c$ as
shown in Fig.~\ref{fig:jc}.
Certainly, the value of $J_c$ will be mainly influenced by
the low-energy part of the host density of states, but
one can make this more precise, trying to answer
the question ``which'' electrons contribute
to the Kondo screening the pseudogap Kondo model.
A rough estimate is provided by the expression for the
critical coupling within the slave-boson mean-field
approximation,
\begin{equation}
\frac{1}{J_c} \propto \int \frac{d\omega}{\omega} \rho_0(\omega) \,,
\end{equation}
with $\rho_0(\omega) \propto |\omega|^r$ being the host density of states.
This integral is logarithmically divergent for $r = 0$,
indicating that the moment is always screened as $T\to 0$,
and that this screening arises from host electrons close
to the Fermi level.
For $r>0$, it is clear that {\em all} conduction electrons
contribute to screening, with a weight proportional to $1/\omega$.
Therefore, an estimate of $J_c$ based on the low-energy part
of $\rho_0$ only may not be appropriate.
We note that a similar expression for $J_c$ can be obtained within
a dynamic large-$N$ approach \cite{OSPG} to the pseudogap
Kondo problem -- there, $\omega$ in the denominator is
replaced by $\omega^{1-\alpha}$, where $\alpha$ describes
the anomalous exponent of the auxiliary fermion propagator
at the critical point (with $\alpha\to 0$ for both $r\to 0$ and
$r\to 1$, but non-zero in between).

\subsection{Behavior above $T_c$}

The numerical calculations in this paper are restricted to
the $d$-wave superconducting state, but we want to give a brief
comment on the normal-state impurity behavior.
The NMR experiments in impurity-doped YBCO \cite{bobroff1,bobroff2}
show enhanced Kondo screening above $T_c$ -- this is
rather natural as the superconducting gap disappears,
the DOS near the Fermi level increases, and the
metallic Kondo effect should be recovered.

Interestingly, no change in the magnetic properties is
seen for underdoped samples when the temperature is tuned
through $T_c$, {\em i.e.}, the NMR susceptibility follows the same
Curie law below and above $T_c$.
Thus, there is no Kondo screening even above $T_c$, which must be
interpreted as a consequence of the celebrated pseudogap.
The fact that no change at all occurs at $T_c$ means that the
depletion of low-energy spectral weight in both the superconducting
and the pseudogap phase is similar -- this appears to support the speculation
that pseudogap and superconducting gap are of the same origin.


\section{Extended magnetic impurities with potential scattering}
\label{sec:4site}

After having discussed general properties of the pseudogap Kondo
model in the context of cuprate superconductors,
we turn to a more detailed spatial modelling of
non-magnetic impurities inducing a magnetic moment.

One important observation from NMR
experiments~\cite{orsay2,julien}
is that these magnetic moments are not localized at
the impurity site, but mainly on the nearest-neighbor copper
sites.
The simplest model for the induced moment is therefore a Kondo model
of the form (\ref{hmag4}),
\begin{equation}
{\cal H}_{\rm mag} = \sum_{\bf R} J_{\bf R} {\bf S} \cdot {\bf s}_{\bf R} \,,
\end{equation}
where the moment is coupled
to the four Cu sites $\bf R$ adjacent to the impurity,
with $J_{\bf R} = J/4$.

An important point is that such a four-site Kondo model (\ref{hmag4})
is not a single-channel model, but instead has four screening
channels associated
with $s$, $p_x$, $p_y$, and $d$-wave-like linear combinations
of the conduction electrons on the four sites.
These channels are not equivalent, and on general grounds
one expects that the low-energy physics is dominated by
the strongest screening channel.
As our NRG cannot deal with a multi-channel model,
we have calculated the critical $J$ values for the
four channels separately, and found that the $d$-wave channel
has the lowest critical $J$ or, equivalently, the highest
$T^\ast$ for fixed $J$.
The physics being dominated by this channel can be
interpreted as the screening cloud having $d$-wave symmetry.
Actually, the same conclusion has been reached in a
slave-boson treatment of the same model \cite{tolya}.
We note that this $d$-wave symmetry is {\em not} directly
related to the $d$-wave symmetry of the superconducting order
parameter of the host -- it is rather a band structure effect
and is completely determined by normal host Green's functions.

We have used such a four-site Kondo model in the $d$-wave channel
to calculate the dashed curves shown in Figs.~\ref{fig:jc}
and \ref{fig:tstardop}.
The effective bath density of states seen by the impurity
in this case is given by
\begin{equation}
\rho_{\rm eff}(\omega) = -\frac{1}{\pi}
{\rm Im}
\sum_{\bf R, R'} \varphi_{\bf R} \varphi_{\bf R'}
{\rm Tr} \left[ G({\bf R}, {\bf R'}, \omega) \frac{1+\tau^z}{2} \right]
\label{rho4site}
\end{equation}
with $\varphi_{\bf R} = +[-]1$ for ${\bf R} - {\bf r}_0 = (\pm 1, 0) [(0,\pm 1)]$,
and $G$ is the $\Psi$ Green's function (\ref{scatgf}) in Nambu notation.
Examples are plotted in Fig.~\ref{fig:dos}b -- is it seen
that the low-energy part close to the Fermi level is
suppressed (actually, it follows an $|\omega|^3$ power law),
whereas the weight at energies around the superconducting
gap is strongly enhanced.
This is easily understood if we consider the $d$-wave-like
coupling of the impurity in momentum space:
the ``form factor'' is $(2 + \cos 2 k_x + \cos 2 k_y - 4 \cos k_x \cos k_y)$,
this expression vanishes along the diagonals of the Brillouin zone, but
has maxima at $(\pi,0)$ and $(0,\pi)$.

A second ingredient for an effective model describing
non-magnetic impurities is a potential scattering term (\ref{hpot}).
This scattering potential is certainly well localized
at the Zn or Li site, however, its bare value $U$ is not
known.
Although certain experiments indicate that Zn behaves like a
strong scatterer, this does not necessarily imply a large
value of the bare scattering potential $U$,
as the discussed Kondo screening can
lead to strong scattering for low temperatures
(note that $T_K$ is much larger in the normal state, and
can reach 150 K or more).
Our results so far have been obtained with zero potential
scattering $U$; we have checked that the results are
practically unchanged for $U$ values up to $t/4$,
and remain qualitatively valid for $U$ up to $2 t$.
For larger values of the bare potential scattering,
the critical $J$ value for the pseudogap Kondo effect
is significantly reduced, due to the increasing density
of states near the Fermi level induced by the nearby
scatterer.

\subsection{Local density of states and STM}

Motivated by the recent observation of impurity-induced
resonances in STM experiments on Zn-doped BSCCO \cite{seamuszn,ali},
we can calculate the local conduction electron density of
states around the impurity.
A possibility first discussed in Ref.~\onlinecite{tolya}
is that the huge peak close to zero bias arises from
the Kondo screening of the Zn-induced moment;
related studies can be found in Ref.~\onlinecite{zhuting}.
[In the STM context,
it is important to distinguish between a screened moment
which acts as a spin-independent resonant scatterer at low temperatures,
and an unscreened moment (as in the case of Ni) which leads
to spin-dependent scattering and causes a splitting of the
impurity-induced peak in the local DOS.]

The local (tunneling) DOS in the presence of both a potential
scatterer and a Kondo impurity is obtained as
\begin{equation}
\rho_{\rm STM}({\bf R},\omega) = -\frac{1}{\pi}
{\rm Im Tr} \left[ \widetilde{G}({\bf R},{\bf R},\omega) \frac{1+\tau^z}{2} \right]
\label{rhostm1}
\end{equation}
where $\widetilde{G}$ is
the full $\Psi$ Green's function
\begin{eqnarray}
&& \widetilde{G}({\bf r},{\bf r}^{\prime},\omega)
= G({\bf r},{\bf r}^{\prime},\omega) \nonumber \\
&+& \sum_{{\bf s},{\bf s}^{\prime}}
\varphi_{\bf s} \varphi_{{\bf s}^{\prime}}
G({\bf r},{\bf s},\omega) \tau^z T(\omega) \tau^z
G({\bf s}^{\prime},{\bf r}^{\prime},\omega)
\label{dos}
\end{eqnarray}
with $T(\omega)$ being the Kondo impurity T matrix,
and the $\bf s$ sites are the neighbors of the impurity at
${\bf r}_0$.

The STM experiments on Zn-doped BSCCO \cite{seamuszn,ali}
show for each impurity a huge peak close to zero bias
(at roughly $-1.5$ meV) at a single location which has to be
identified with the impurity site.
This peak signal decays rapidly when moving away from the impurity,
in addition it shows a checkerboard-like spatial modulation,
{\em i.e.}, it does not occur on the nearest neighbors of
the impurity, but on the next-nearest neighbors and so on.
This spatial shape is hard to explain with the assumption of
strong potential scattering on the impurity site, because
a large bare $U$ would expel all electrons there and lead
to a very small tunneling DOS exactly {\em at} the impurity
site.

When comparing the results of microscopic calculations
with STM data, one has to think
about the actual tunneling path of electrons between the STM tip
and the CuO$_2$ plane.
BSCCO crystal are cleaved in a way that a BiO layer forms
the sample surface, and the electrons presumably tunnel via
this BiO plane into the CuO$_2$ plane of interest.
The nature of this tunneling path is not known.
It has been proposed recently~\cite{filter1,filter2}
that tunneling from the STM tip into a certain
Bi orbital actually probes the electrons on the {\em neighboring}
Cu 3d orbitals, leading to a strongly momentum-dependent
tunneling matrix element.
This so-called filter effect resolves the discrepancy between
the observed spatial shape on the Zn resonance and the result
expected from strong potential scattering.
More precisely, the tunneling matrix elements proposed in
Ref.~\onlinecite{filter2} have a $d$-wave shape, such that
the STM signal is given by the modified local density of
states
\begin{eqnarray}
\rho_{\rm STM}({\bf R},\omega) &=& -\frac{1}{\pi}
{\rm Im}
\sum_{\bf R', R''} \bar{\varphi}_{\bf R'} \bar{\varphi}_{\bf R''} \nonumber\\
&& {\rm Tr}
\left[\widetilde{G}({\bf R'}, {\bf R''}, \omega) \frac{1+\tau^z}{2} \right]
\end{eqnarray}
replacing (\ref{rhostm1}),
with $\bar{\varphi}_{\bf R'} = +[-]1$ for ${\bf R'} - {\bf R} = (\pm 1, 0) [(0,\pm 1)]$.
[Note the similarity to the effective DOS seen by the four-site impurity,
eq.~(\ref{rho4site}).]
At present, it is not clear whether such filtering actually occurs,
and further experiments are needed to clarify the tunneling path
into the CuO$_2$ plane.

\begin{figure}[t]
\epsfxsize=3.4in
\centerline{\epsffile{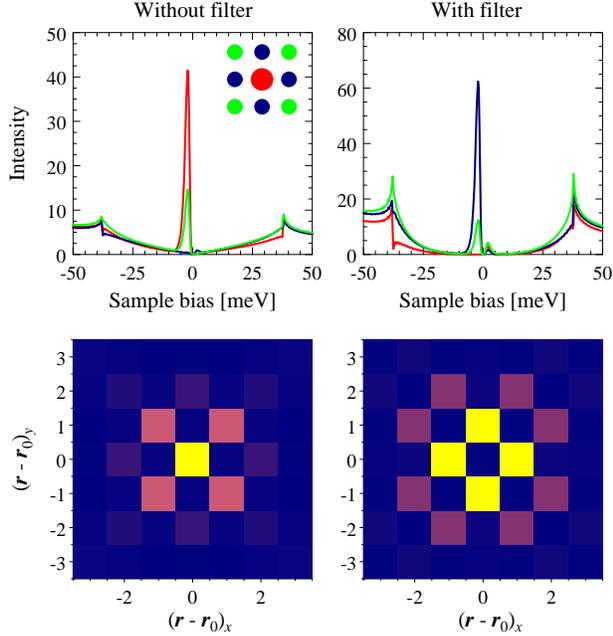}}
\caption{
Calculated tunneling density of states for the four-site
Kondo impurity model at 15\% hole doping with a
realistic band structure ($t=-0.15$ eV, $t^{\prime} = -t/4$,
$t^{\prime\prime} = t/12$), $\Delta_0 = 0.04$ eV, and
$\mu = -0.14$ eV.
The Kondo coupling is $J=0.09$ eV, the potential
scattering $U=0$.
Top: Local DOS vs. energy for the
impurity site (red) and the nearest (blue) and second (green)
neighbor sites.
Bottom: Spatial dependence of the local DOS at $\omega = -2$ meV.
Left: Local DOS in the CuO$_2$ plane.
Right: Local DOS after applying the filter effect
proposed in Ref.~\protect\onlinecite{filter2}.
}
\label{fig:stm1}
\end{figure}

The paper Ref.~\onlinecite{tolya} ignored this filter effect
(as well as the possibility of strong potential scattering
at the impurity site) - we will show here that
the results of Ref.~\onlinecite{tolya} regarding the spatial
shape of the impurity resonance are essentially unchanged
for weak to moderate potential scattering, but
change qualitatively for strong potential scattering.
In the following, we will display the results of our calculations
both with and without accounting for this possible filter
effect~\cite{filter2}.

Fig.~\ref{fig:stm1} shows the calculated tunneling spectra for
a four-site Kondo impurity with zero (or weak) potential
scattering.
These data are similar to Fig. 1 of Ref.~\onlinecite{tolya},
but here the impurity properties are calculated using NRG
removing some artifacts of the slave-boson method,
namely too large values of the critical coupling $J_c$,
too large Kondo temperatures for $J>J_c$,
and too sharp peaks in the T matrix.
It is clearly seen that only the data in the left panel, {\em i.e.},
without the filter effect, are consistent with
the experimental observation~\cite{seamuszn}, in that they show a
large peak {\em at} the impurity site.
Remarkably, the so-called coherence peaks in the tunneling DOS
at $\pm \Delta_0$ are almost completely suppressed near the impurity
(cf. Fig.~\ref{fig:dos} for the bulk DOS) due to the
interference between the bulk Green's functions and the impurity T
matrix,
although the superconducting order parameter is {\em not} changed
(we did not account for gap relaxation).
Therefore the interpretation of suppressed coherence peaks
in terms of locally suppressed superconductivity has to be used
with care.

\begin{figure}[t]
\epsfxsize=3.4in
\centerline{\epsffile{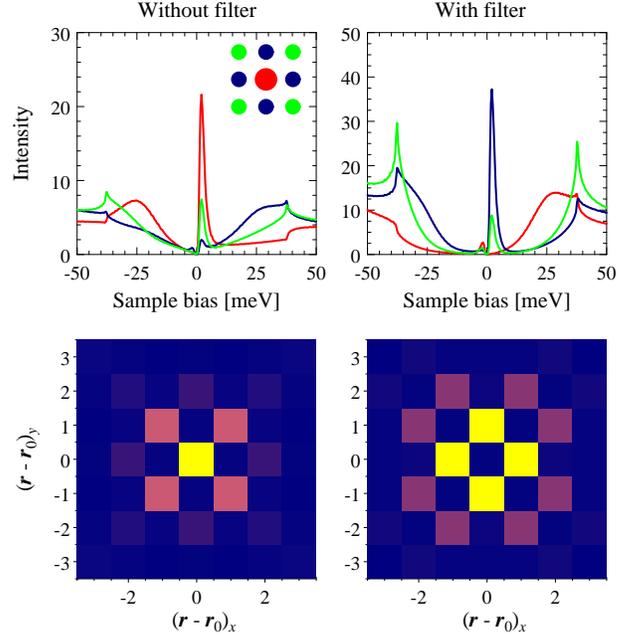}}
\caption{
Same as Fig.~\protect\ref{fig:stm1}, but with
potential scattering $U=|t|=0.15$ eV.
Here, $J = 0.065$ eV.
The lower panel shows the local DOS at $\omega = +2$ meV.
}
\label{fig:stm2}
\end{figure}

Switching on moderate potential scattering on the impurity
site does not qualitatively modify the picture, as shown in
Fig.~\ref{fig:stm2}.
Note that for the chosen band structure and potential scattering value
the global particle-hole asymmetry has changed its sign,
therefore the Kondo peak appears at the opposite side of the
Fermi level compared to Fig.~\ref{fig:stm1}.
(This should not be seen as a contradiction to the experiment,
as the tight-binding band structure is a rough approximation
to the real system, and the sign of the overall particle-hole
asymmetry is not known.)

\begin{figure}[t]
\epsfxsize=3.4in
\centerline{\epsffile{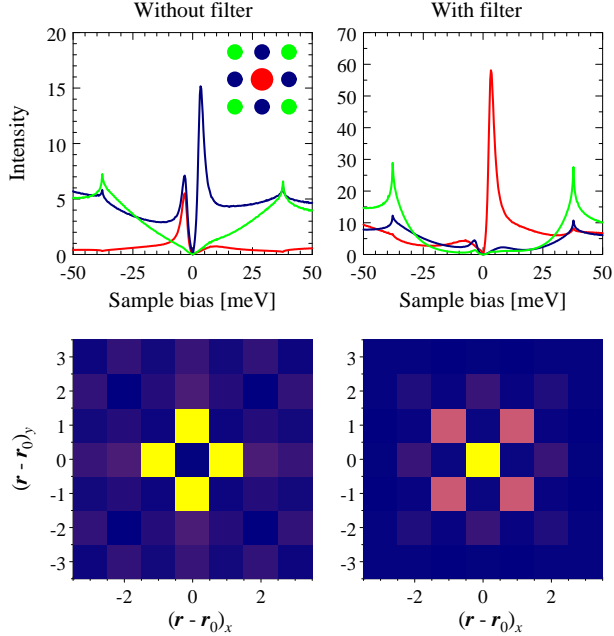}}
\caption{
Same as Fig.~\protect\ref{fig:stm1}, but with
potential scattering $U=4|t|=0.6$ eV.
Here, $J = 0.04$ eV.
The lower panel shows the local DOS at $\omega = +3$ meV.
}
\label{fig:stm3}
\end{figure}

For huge values of potential scattering, the Kondo effect
is heavily influenced in the present simple model.
Due to the large scattering-induced DOS on the neighboring
sites of the impurity the critical Kondo coupling is
strongly reduced.
Correspondingly the Kondo temperature
for reasonable values of $J$ ({\em e.g.}, 50-100 meV)
is increased to 100 K or more, which is in disagreement
with the NMR observation \cite{bobroff2}.
Therefore, we consider such a parameter combination
unlikely to describe the experimental situation.
Nevertheless, we have calculated the STM spectrum for
such a case, shown in Fig.~\ref{fig:stm3}.
We have chosen a smaller value of $J$, to obtain a
Kondo temperature comparable to the experimental value.
It is seen that a double peak structure can occur.
There are strong interence effects between the Kondo peak
in the impurity T matrix and the potential scattering peak
in the host DOS, with the result that the local (tunneling)
DOS is completely dominated by the potential scattering
peak.
The spatial shape of the resulting pattern is more
compatible with the experiment after the filter
effect is taken into account.

\begin{figure}[t]
\epsfxsize=3.4in
\centerline{\epsffile{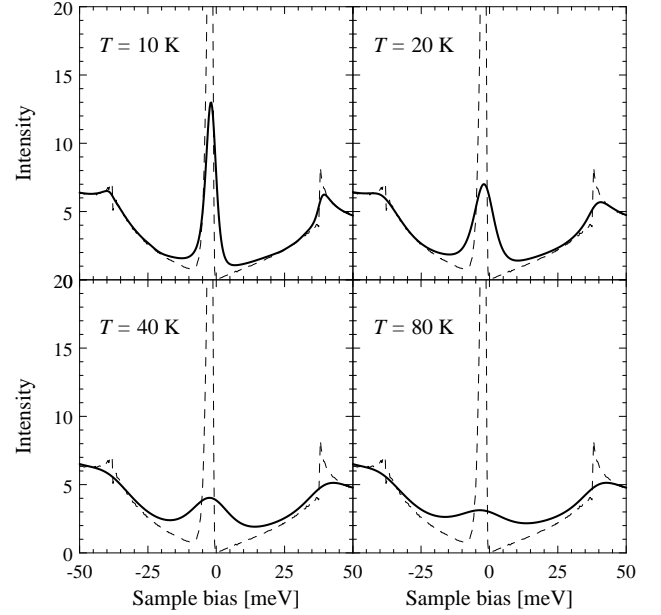}}
\caption{
Temperature evolution of the STM spectrum
shown in Fig.~\protect\ref{fig:stm1},
$T^\ast\!\approx\!20$ K,
calculated with finite-temperature NRG
under the assumption of a $T$-independent
host DOS.
Shown is the tunneling DOS at the impurity site
without applying the filter effect;
the dashed line is the $T=0$ result
from Fig.~\protect\ref{fig:stm1}.
The thermal broadening of the signal due to
the Fermi functions in both tip and sample has been
taken into account.
}
\label{fig:stm4}
\end{figure}

Last not least, in Fig.~\ref{fig:stm4}
we show the temperature variation of
the STM spectra for the situation in Fig.~\ref{fig:stm1}.
Besides the simple thermal broadening due to Fermi functions
in tip and sample, there is additional broadening and
weight loss the Kondo peak as shown in Fig.~\ref{fig:gimptg0}.
Note that we have assumed a temperature-independent
host DOS, which is approximately justified for temperatures
up to $0.7\,T_c$.
The broadening and the loss of spectral
weight occur rather gradually, and the peak will be
seen also for temperature well above $T_K$.
If $T^\ast \approx T_K\approx 20$ K in optimally doped BSCCO
then the broadened Kondo peak will survive almost up to
$T_c$.
Near $T_c$ the gap will close, $T_K$ will increase \cite{bobroff2},
and the Kondo-induced peak might turn into a dip or Fano-like lineshape
as seen for Kondo impurities in metals \cite{stmmetal}.
However, this signal will be broad and perhaps hard to observe.

Closing this section, we mention that there are
subtleties in interpreting the NMR experiments~\cite{orsay2,julien}
in terms of a moment located on four sites
(which lead to our simple effective model).
The NMR experiments measure of course the fluctuating fields
in the {\em full} system, whereas our model specifies the
location of the {\em bare} moment.
For a fair comparison one would have to calculate
local NMR spectra after having accounted for the
interaction of the moment with the host fermions (and
possibly also with the antiferromagnetic spin fluctuations).
This is possible in principle, but is left for future studies.


\section{Conclusions}
\label{sec:concl}

In this paper, we have studied the dynamics of magnetic moments
in high-temperature superconductors.
We have focussed on several aspects of the pseudogap Kondo model
which describes the interaction of a localized spin with
fermionic quasiparticles which obey a linearly vanishing
density of states near the Fermi level.
This Kondo problem shows a non-trivial boundary quantum phase
transition as function of the Kondo coupling (or the size
of the $d$-wave gap); and we have argued that the
quantum-critical behavior has consequences for
susceptibility, NMR, and STM measurements in the cuprate
superconductors.
In particular, the cross-over (``Kondo'') temperature $T^\ast$,
as measured by NMR experiments, can acquire a strong doping
dependence, due to the depletion of low-energy spectral
weight with decreasing doping;
and we have proposed that the observed vanishing of the
Kondo temperature can be explained in terms of the
quantum phase transition in the pseudogap Kondo model.
The impurity T matrix which is observable via STM shows a
large peak at a finite energy which coincides with the
cross-over temperature $T^\ast$.

Our calculated STM spectra are consistent with the experimental
observation~\cite{seamuszn} for small to moderate potential
scattering values, and without taking into account the
filter effect proposed in Ref.~\onlinecite{filter2}.
Within our simple effective model,
larger potential scattering values lead to Kondo temperatures
at variance with the NMR results~\cite{bobroff2}.
Of course, we cannot exclude that the spatial distribution
of the Zn-induced moment is different from what we have
assumed, which would of course strongly affect the spatial
shape of the STM resonance
(but only weakly change the magnetic properties).
One key experiment to decide whether the STM peak is
of Kondo origin would be to measure its temperature
evolution.
Here the characteristic quantity is the spectral weight
associated with the local Zn resonance.
However, the weight loss with increasing $T$ is rather
slow (Figs.~\ref{fig:gimptg0} and \ref{fig:peakweight}),
therefore high-quality measurements on impurities with
small $T_K$ ({\em i.e.}, small bias of the impurity resonance
peak) are required.

The present NRG calculations for the pseudogap Kondo model
provide more reliable results
than the common slave-boson mean-field approximation.
This holds both for the qualitative behavior of the cross-over
scale $T^\ast$ near the transition, and the quantitative
results for the critical Kondo coupling and actual Kondo
temperatures for parameters relevant for cuprates.

In the present work, we have neglected both the relaxation
of the local superconducting order parameter around each impurity
as well as interactions between the impurities.
The latter are exponentially small in the impurity density and
are therefore unimportant for small impurity concentration
(in the experimentally accessible temperature regime).
The gap relaxation will mainly lead to a small residual
low-energy DOS near the impurity, which causes
Kondo screening at low enough temperatures even in the
underdoped regime.
However, the induced Kondo temperature is exponentially
small in the residual DOS, and can be safely neglected.

Summarizing,
our studies support that the NMR observations of
Curie-Weiss-like behavior associated with
Li or Zn impurities in high-$T_c$ compounds
can be described by the Kondo screening of
the impurity-induced moments.
The peak in the local DOS in Zn-doped BSCCO is likely of
the same origin, but a number of experimental as well as theoretical
issues remain to be resolved.
We emphasize that the given numbers for Kondo couplings
and characteristic temperatures can be viewed as rough estimates
only, since the precise form of a Hamiltonian describing the
impurity-induced moment is not known.

Further theoretical studies should address the spatial dependence of
NMR spectra ({\em e.g.}, on the Y sites near the impurity),
the additional coupling to antiferromagnetic fluctuations,
as well as implications for transport measurements.


\acknowledgements

We thank H. Alloul, A. V. Balatsky,
J. Bobroff, S. Davis, M. Flatt\'{e}, K. Ingersent,
D. E. Logan, A. Matsuda, T. Pruschke, and N.-C. Yeh
for helpful discussions,
and especially S. Sachdev and A. Polkovnikov for
collaboration on related work as well as numerous
invaluable discussions.
This research was supported by the DFG through SFB 484.



\begin{references}
\bibitem{phen}
This phenomenological view implies that it is possible
to separate the formation of impurity moments from their
interaction with the low-energy bulk excitations.
As the moment formation is a high-energy phenomenon,
this assumption appears justified regardless whether the
moment is eventually screened or not.
\bibitem{ph} If the host band structure is particle-hole symmetric
(not the case at optimal doping), the potential
scattering ($U$ in $H_{\rm imp}$) has to be near
infinity
to obtain a low energy quasi-bound state, but in general, some finite
and non-universal value of $U$ is required.
\end{references}
\end{document}